\documentclass[10pt]{iopart}
\usepackage{graphics,amsfonts}
\begin{document}
\title{Monte-Carlo and Bayesian techniques in gravitational wave burst data analysis}
\author{Antony C Searle$^1$
}
\address{$^1$ LIGO - California Institute of Technology, Pasadena, CA 91125}

\ead{$^1$\mailto{acsearle@ligo.caltech.edu}
}
\newcommand{\gursel}{G\"{u}rsel}
\begin{abstract}
Monte-Carlo simulations are used in the gravitational wave burst detection community to demonstrate and compare the properties of different search techniques.  We note that every Monte-Carlo simulation has a corresponding optimal search technique according to both the non-Bayesian Neyman-Pearson criterion and the Bayesian approach, and that this optimal search technique is the Bayesian statistic.  When practical, we recommend deriving the optimal statistic for a credible Monte-Carlo simulation, rather than testing ad hoc statistics against that simulation.
\end{abstract}
\submitto{\CQG}
\maketitle
\section{Introduction}
A number of different statistics \cite{GuTi:89,Tinto96,FlHu:98b,AnBrCrFl:01,KlMoRaMi:05,AEI05,KlMoRaMi:06,MoRaKlMi:06,Ra:06,Ch_etal:06,searle:08} have been proposed to detect bursts of gravitational waves with the global network of gravitational wave interferometers.  With the exception of \cite{FlHu:98b,AnBrCrFl:01,searle:08} these have been derived in a frequentist or \emph{ad hoc} framework.  The tool used by the community to demonstrate and compare their performance is the well-known Monte-Carlo simulation.  The simulations are intuitive and easy to implement \cite{nr}.  To perform a Monte-Carlo simulation requires generating many sample observations from two populations, one consisting of background noise only and the other background noise with gravitational wave signals added (or ``injected'', in community terminology).

The author, with Sutton, Tinto and Woan, has elsewhere \cite{searle:08} proposed a Bayesian statistic for the detection of gravitational wave bursts.  Conventionally, the next step would be to test its performance using a Monte-Carlo simulation (and such a study is forthcoming).  Yet the notion of testing a Bayesian gravitational wave burst search with a Monte-Carlo simulation raises some novel issues.  Adopting concrete models for noise and signal is required by both Monte-Carlo simulations and a Bayesian analysis, and moreover the Bayesian statistic so defined can be shown to be the best possible statistic for the corresponding Monte-Carlo simulation \emph{under the non-Bayesian Neyman-Pearson criterion}.  Performing a Monte-Carlo simulation will not determine if the corresponding Bayesian statistic performs better than other statistics; it can only quantify how much better its performance is.

The relationship between Monte-Carlo simulations and Bayesian analyses is due to their common assertion that, despite our incomplete knowledge of the population of gravitational wave bursts, we can learn something by proceeding as if gravitational wave bursts had some particular distribution.  In the Bayesian case, this distribution is the prior plausibility distribution for the signal model.  In the Monte-Carlo case, this is the probability distribution we sample from to produce the signal population.  In both cases, there is no right choice, but there are many wrong choices that contradict our physical knowledge, and ways to prefer some choices over others on grounds of including more physical knowledge.  The credibility we attach to a Monte-Carlo simulation or a Bayesian analysis hinges on if we are satisfied that the signal (and noise) models involved adequately represents our state of knowledge about the universe.

In \S\ref{sec:mc} we formally review Monte-Carlo simulations as typically conducted by the bursts community.  In \S\ref{sec:np} we note the Neyman-Pearson criterion, and its relationship to Monte-Carlo simulations.  In \S\ref{sec:models} we look at how statistics and signal models imply each other and what this means for papers that use two conflicting definitions.  In \S\ref{sec:bayes} we discuss the equivalence between the Monte-Carlo simulation signal model and the Bayesian signal prior.  In \S\ref{sec:limitations} we note some conditions that can restrict the applicability of our conclusions.

\section{Analysis}

\subsection{Monte-Carlo simulations\label{sec:mc}}

In the problem of gravitational wave burst detection, we wish to classify an observation $\mathbf{x}$ as either \emph{signal} or \emph{noise}.  A deterministic rule will classify every possible observation as either signal or noise, partitioning the space of observations into two regions $R_\mathrm{signal}$ and $R_\mathrm{noise}$.  The classification will be imperfect, because it is possible for any particular observation to arise from signal or noise (for example, an observation could be due to a gravitational wave, or it could be due to a large noise excursion.)  Misclassifying a signal as noise is a \emph{false dismissal}; misclassifying noise as a signal is a \emph{false alarm}.  Different classification rules will partition the space of observations into different regions, and make different mistakes for different kinds of observation.  To compare classification rules, we can perform a Monte-Carlo simulation.

First consider the noise hypothesis $H_\mathrm{noise}$.  To perform the Monte-Carlo simulation, we need many observations $\mathbf{x}$ corresponding to many different realizations of the noise.  A simple Monte-Carlo simulation might create many white noise time-series, each with $n$ time samples.  In this case, we are sampling the distribution
\begin{eqnarray}
p(\mathbf{x}|H_\mathrm{noise})=(2\pi)^{-\frac{n}{2}}\exp(-\frac{1}{2}\mathbf{x}^T\mathbf{x}).
\end{eqnarray}
A realistic simulation might instead use previous observations of the gravitational wave observatory.  In this case, we are still sampling a distribution $p(\mathbf{x}|H_\mathrm{noise})$, but we do not have an explicit form for the distribution.

We can compute the \emph{false alarm probability} $p_F$ (the fraction of noise realizations misclassified as signals) for the classification rule by evaluating the integral
\begin{eqnarray}
p_F=\int_{R_\mathrm{signal}}p(\mathbf{x}|H_\mathrm{noise})\mathrm{d}\mathbf{x}.
\end{eqnarray}
Note that we can only do this once we have chosen a distribution $p(\mathbf{x}|H_\mathrm{noise})$; it is not an intrinsic property of the classification rule.  This is because we are measuring the portion of the specified distribution inside region $R_\mathrm{signal}$.

The integral evaluates a potentially implicitly defined distribution over a potentially implicitly defined region.  The technique of Monte-Carlo integration with \emph{importance sampling} fits exactly this situation \cite{nr}.  Instead of sampling the space of observations uniformly, we sample it with distribution $p(\mathbf{x}|H_\mathrm{noise})$.  The integral then becomes
\begin{eqnarray}
p_F&=&\int_{\mathbb{R}^n}\left\{
\begin{array}{cl}
1 & \mathrm{if\ }\mathbf{x}\in R_\mathrm{signal}\\
0 & \mathrm{otherwise}
\end{array}
\right\}(p(\mathbf{x}|H_\mathrm{noise})\mathrm{d}\mathbf{x})\\
&=&\left\langle\frac{1}{N}\sum_{i=1}^N\left\{
\begin{array}{cl}
1 & \mathrm{if\ }\mathbf{x_i}\in R_\mathrm{signal}\\
0 & \mathrm{otherwise}
\end{array}
\right\}\right\rangle
\end{eqnarray}
for $N$ sample observations $\mathbf{x}_i$ drawn from $p(\mathbf{x}|H_\mathrm{noise})$.  This is how the intuitively obvious counting strategy, where we draw many samples and count how many lie in the region, is related to the formal definition of the false alarm fraction.

Now consider the hypothesis $H_\mathrm{signal}$.  To perform a Monte-Carlo simulation, we need many observations $\mathbf{x}_i$ drawn from a distribution $p(\mathbf{x}|H_\mathrm{signal})$.  In the literature, this distribution is typically implicitly defined by a procedure like the one we now describe:

A waveform $\mathbf{h}$ is drawn from some distribution $p(\mathbf{h})$.  Often the distribution is a set of a few discrete waveforms $\mathbf{h}_i$:
\begin{eqnarray}
p(\mathbf{h})&=&\frac{1}{N_h}\sum_{i=1}^{N_h}\delta(\mathbf{h}-\mathbf{h}_i).
\end{eqnarray}
An amplitude $\sigma$ for the injection is drawn from a distribution $p(\sigma)$, again often from a few discrete choices:
\begin{eqnarray}
p(\sigma)&=&\frac{1}{N_\sigma}\sum_{i=1}^{N_\sigma}\delta(\sigma-\sigma_i).
\end{eqnarray}
A direction on the sky $(\theta,\phi)$ is drawn from a distribution $p(\theta,\phi)$, usually representing a uniform distribution on the sky:
\begin{eqnarray}
p(\theta,\phi)&=&p(\theta)p(\phi)\\
&=&\left\{
\begin{array}{cl}
\sin\theta&0<\theta<\pi\\
0&\mathrm{otherwise}
\end{array}
\right\}\left\{
\begin{array}{cl}
\frac{1}{2\pi}&-\pi<\phi<\pi\\
0&\mathrm{otherwise}
\end{array}
\right\}.
\end{eqnarray}
The direction $(\theta,\phi)$ allows the computation of the linear response of the network to strain, a matrix $\mathbf{F}(\theta,\phi)$ including antenna patterns and time-delays.  Finally, background noise $\mathbf{e}$ is drawn from the distribution $p(\mathbf{e}|H_\mathrm{signal})=p(\mathbf{x}|H_\mathrm{noise})$.

The sample of $p(\mathbf{x}|H_\mathrm{signal})$ is then constructed as by adding the strain, transformed by the response, and scaled by the amplitude, to the background noise:
\begin{eqnarray}
\mathbf{x}&=&\sigma\mathbf{F}\cdot\mathbf{h}+\mathbf{e}
\end{eqnarray}
or, expressed as a distribution, drawing $\mathbf{x}$ from
\begin{eqnarray}
p(\mathbf{x}|\mathbf{h},\sigma,\theta,\phi,\mathbf{e})&=&\delta(\mathbf{x}-\sigma\mathbf{F}\cdot\mathbf{h}-\mathbf{e})
\end{eqnarray}
As we know the distributions of all the components, we can compute an explicit expression for the $H_\mathrm{signal}$ distribution
\begin{eqnarray}
p(\mathbf{x}|H_\mathrm{signal})&=&
\int_{\mathbb{R}^m}p(\mathbf{x}|\mathbf{h},\sigma,\theta,\phi,\mathbf{e})p(\mathbf{h})p(\sigma)p(\theta,\phi)p(\mathbf{e})\rmd\mathbf{h}\,\rmd\sigma\,\rmd\theta\,\rmd\phi\,\rmd\mathbf{e}.\nonumber\\
&&\
\end{eqnarray}
This integral explores all possible combinations of parameters, weights them by their joint probability of occurring, and then assigns that probability to the observation they will result in.  The resulting expression for the distribution is unwieldy, but samples can be readily drawn from it by following the procedure above.  (The signal model so described, with a handful of waveforms at a handful of fixed distances, is far from physical.)

The \emph{detection probability} $p_D$ (the fraction of realizations of the signal hypothesis correctly classified as signals) can be evaluated like the false alarm probability:
\begin{eqnarray}
p_D&=&\int_{R_\mathrm{signal}}p(\mathbf{x}|H_\mathrm{signal})\mathrm{d}\mathbf{x}\\
&=&\int_{\mathbb{R}^n}\left\{
\begin{array}{cl}
1 & \mathrm{if\ }\mathbf{x}\in R_\mathrm{signal}\\
0 & \mathrm{otherwise}
\end{array}
\right\}(p(\mathbf{x}|H_\mathrm{signal})\mathrm{d}\mathbf{x})\\
&=&\left\langle\frac{1}{N}\sum_{i=1}^N\left\{
\begin{array}{cl}
1 & \mathrm{if\ }\mathbf{x_i}\in R_\mathrm{signal}\\
0 & \mathrm{otherwise}
\end{array}
\right\}\right\rangle
\end{eqnarray}
where the observations $\mathbf{x}_i$ are in this case drawn from the signal distribution $p(\mathbf{x}|H_\mathrm{signal})$.  Again, the detection fraction computed depends on the signal distribution chosen; it is not an intrinsic property of the classification rule.

The false alarm and detection probabilities tell us something about the performance of the classification rule.  Small false alarm probabilities and large detection probabilities are desirable; they mean the classification rule will make few mistakes, but this quality is contingent on the signal and noise hypotheses accurately representing the physical system we want to apply the rule to.  In gravitational wave detection, we wish to set the false alarm probability to some acceptable level, such as one false alarm in one hundred years of observation time.  For this fixed false alarm probability, we want to choose the classification rule that maximizes the detection probability; this is the \emph{Neyman-Pearson criterion} we explore in the next section.

\subsection{The optimal statistic\label{sec:np}}

One way to form a classification rule is to compute a function (or \emph{statistic}) of the observation $f(\mathbf{x})$, and compare it against a threshold $\lambda$.  Then
\begin{eqnarray}
\mathbf{x}\in R_\mathrm{signal}&\Leftrightarrow&f(\mathbf{x})>\lambda\\
\mathbf{x}\in R_\mathrm{noise}&\Leftrightarrow&f(\mathbf{x})\leq\lambda
\end{eqnarray}
An example statistic is the (dimensionless quantity proportional to) energy, $f(\mathbf{x})=\mathbf{x}^T\mathbf{x}$.

If we vary the threshold $\lambda$ we can produce a family of decision rules, and the detection $p_D$ and false alarm $p_F$ probabilities become functions of $\lambda$, $p_D(\lambda)$ and $p_F(\lambda)$.  The two-dimensional curve parameterized by $\lambda$, $(p_F(\lambda),p_D(\lambda))$ is called the \emph{receiver operating characteristic}.  An approximation to the receiver operating characteristic curve can be computed as cheaply as a single Monte-Carlo evaluation for $p_F$ and $p_D$, by recording $f(\mathbf{x})$ for each realization rather than only noting if it exceeds $\lambda$.  Approximations to $p_F(\lambda)$ and $p_F(\lambda)$ can then be cheaply computed by counting the fraction of the precomputed values above the threshold, but the estimates so obtained are not independent for different values of $\lambda$.

The frequentist Neyman-Pearson criterion \cite{helstrom} says that we should choose our classification rule to maximize the detection probability $p_D$ for a given false alarm probability $p_F$.  The maximization is over the space of all possible classification rules.  There exists an explicit solution to this optimization problem given by the \emph{Neyman-Pearson lemma}:
\begin{eqnarray}
\Lambda(\mathbf{x})&=&\frac{p(\mathbf{x}|H_\mathrm{signal})}{p(\mathbf{x}|H_\mathrm{noise})}>\lambda
\end{eqnarray}
for $\lambda$ yielding the desired $p_F(\lambda)$.  $\Lambda$ is called the \emph{likelihood ratio}.  It is important to note that this optimality is for the particular signal model adopted for the Monte-Carlo simulation.  It is not a claim for all signal models.

It is easy to see why this choice is optimal.  Break $\mathbb{R}^n$ into infinitesimal pieces $\rmd\mathbf{x}$.  We want to incrementally construct $R_\mathrm{signal}$ from these pieces.  The best piece, according to the Neyman-Pearson criterion, brings the most signal realizations for a given number of noise realizations.  This is the piece with the maximal $(p(\mathbf{x}|H_\mathrm{signal})\rmd\mathbf{x})/(p(\mathbf{x}|H_\mathrm{noise})\rmd\mathbf{x})=\Lambda(\mathbf{x})$. The next best piece has the next highest $\Lambda(\mathbf{x})$ and so on.  We take pieces in this fashion until $p_F$ for their union reaches the desired value, which is the same as taking all pieces with $\Lambda(\mathbf{x})>\lambda$.

The claim that $\Lambda(\mathbf{x})$ is the optimal statistic may appear to be in conflict with the non-existence of a uniformly most powerful test in this situation \cite{helstrom}.  The concept of a uniformly most powerful test refers to \emph{parameterized} hypotheses, and in this scenario it means that if we seek to decide between a \emph{specific} signal $H_{\mathbf{h},\theta,\phi,\sigma}$ and $H_\mathrm{noise}$, the most powerful test (i.e. $R_{\mathbf{h},\theta,\phi,\sigma}$) will vary (not be \emph{uniform}) depending on which \emph{specific} signal $(\mathbf{h},\theta,\phi,\sigma)$ we consider.  This is not the case for the Monte-Carlo simulation, where we do not look for a specific signal, but instead for \emph{any} signal from some specific distribution.  The non-existence of the uniformly most powerful test is a special case of the observation that the optimal statistic depends on the signal model we adopt.

\subsection{Implicit and explicit signal models\label{sec:models}}

We can infer the optimal statistic for a given signal and noise model, but we can also work backwards from an arbitrary statistic to infer the signal and noise distributions it is the optimal statistic for.  In general, the identification cannot be made uniquely; arbitrary strictly increasing functions of one statistic yield another statistic with identical (though differently parameterized) classification regions; nor can the likelihood ratio uniquely define its numerator and denominator distributions.  However, when we have access to the derivation of the statistic, as in \cite{GuTi:89,Tinto96,KlMoRaMi:05,KlMoRaMi:06,MoRaKlMi:06,Ra:06}, where the noise model is explicitly asserted and the relationship of the statistic to probability is known (typically the logarithm of a probability ratio), we can readily extract a ``natural'' member of the set of statistics that make the same classifications.  In \cite{searle:08} this is done for several statistics used in the community and it is found that they implicitly correspond to unphysical or even improper signal distributions: a population of gravitational waves whose energies at Earth are infinite, infinitesimal or have some particular value and are anisotropically distributed across the sky.

When a proposed statistic is tested against a Monte-Carlo simulation, the paper effectively proposes two physical models, one implicit in the statistic and one more explicit in the simulation.  When these models differ (as they do in the papers cited above) the paper contradicts itself, and the test intended to demonstrate the statistics efficacy instead suggests a different statistic. This optimal statistic makes its gains by fully adapting to the simulation; if the simulation uses only a handful of waveforms and distances it can even outperform a matched filter bank (which searches all distances).  We might view the optimal statistic as ``cheating'', but the fault lies with the unphysicality of the simulation; to even form the notion of cheating is to admit that the Monte-Carlo simulation we have used is not compatible even with our incomplete state of knowledge about the physical signal population.  More importantly, testing with such a simulation does not conclusively demonstrate that the proposed statistic is itself not cheating in some fashion.  An unphysical Monte-Carlo simulation can give us no assurances about a ``black box'' statistic.

A better path is simply to explicitly propose a physically credible model and then derive the optimal statistic for it.

\subsection{Bayesian interpretation\label{sec:bayes}}

The Bayesian \cite{gregory} solution to the classification problem is
\begin{eqnarray}
\frac{p(H_\mathrm{signal}|\mathbf{x})}
{p(H_\mathrm{noise}|\mathbf{x})}
&>&1
\end{eqnarray}
where the posterior (``after'' we consider the observation) plausibility ratio is defined as the product of the \emph{Bayes factor} and the prior (``before'' we consider the observation) plausibility ratio:
\begin{eqnarray}
\frac{p(H_\mathrm{signal}|\mathbf{x})}
{p(H_\mathrm{noise}|\mathbf{x})}
&=&
\frac{p(\mathbf{x}|H_\mathrm{signal})}
{p(\mathbf{x}|H_\mathrm{noise})}
\frac{p(H_\mathrm{signal})}
{p(H_\mathrm{noise})}
\end{eqnarray}
The Bayes factor is the ratio of how well each hypothesis predicted the observation $\mathbf{x}$; it tells us how much we should change our initial  opinion, represented by the prior plausibility ratio, into our final opinion, represented by the posterior plausibility ratio.   It is computed by forming marginalization integrals over \emph{nuisance} parameters and their prior distributions
\begin{eqnarray}
\frac{p(\mathbf{x}|H_\mathrm{signal})}
{p(\mathbf{x}|H_\mathrm{noise})}
&=&
\frac{
\int p(\mathbf{x}|\sigma,\theta,\phi,\mathbf{h},\mathbf{e},H_\mathrm{signal})
p(\sigma)p(\theta,\phi)p(\mathbf{h})p(\mathbf{e})\rmd\sigma\,\rmd\theta\,\rmd\phi\,\rmd\mathbf{h}\,\rmd\mathbf{e}
}
{
p(\mathbf{x}|H_\mathrm{noise})
}\nonumber\\
&&\
\end{eqnarray}
The mathematics involved is identical to the Monte-Carlo simulation.  The Monte-Carlo simulation's parameter probability distributions are the Bayesian analyses' prior plausibility distributions.  The Monte-Carlo procedure of sampling them is the same as Bayesian marginalization over them.  The Bayes factor is the likelihood ratio. The likelihood threshold $\lambda$ is the inverse of the Bayesian prior plausibility ratio $p(H_\mathrm{signal})/p(H_\mathrm{noise})$.  The Bayesian solution to the classification problem is the Neyman-Pearson optimal statistic for the Monte-Carlo simulation.

The close relationship between Monte-Carlo simulations and Bayesian analysis is simply explained.  Both assert that, despite incomplete knowledge of the physical signal model, we can learn something by proceeding with a signal model that is consistent with what we do know (and appropriately non-committal with respect to what we don't know).

The notion of testing the Bayesian statistic with a Monte-Carlo simulation clarifies the inconsistency implicit in Monte-Carlo testing \emph{ad hoc} statistics.  If we test the Bayesian statistic with a different signal model, we must mean that we believe that signal model to be more plausible than the one used to derive the Bayesian statistic, in which case we should have derived the Bayesian statistic from it in the first place.  (Testing with different models lets us measure relative ``robustness'', but the proper way to be robust is to use a signal model which is the appropriately weighted average of all the different signal models we hope to be robust over.)  Of course, performing a Monte-Carlo simulation can still be valuable, allowing us to evaluate the receiver-operating characteristic, validate the implementation, or compare performance to \emph{ad hoc} statistics, but by definition it will not shed further light on \emph{if} the optimal statistic is or is not better than other statistics.

\subsection{Limitations\label{sec:limitations}}

None of the above suggests that the statistics in \cite{GuTi:89,Tinto96,KlMoRaMi:05,KlMoRaMi:06,MoRaKlMi:06,Ra:06} do not work, or even that they do not work well, but only that they do not work as well as is possible.  Fortunately, the simplicity of the signal models employed in these searches means that they are likely to be sufficiently sampled by the limited Monte-Carlo simulations they are tested with (i.e. unlikely to ``cheat'').  The main drawback is the inability to produce a single-number event rate for a disparate and properly spatially distributed source population, a limitation that is not seen as major given the community's focus on computing detection ranges for given source energies.

Practically speaking, it is only possible to derive an optimal statistic when the signal and noise models are known.  In the case of the real instruments, the noise model is not fully known, and any Monte-Carlo simulation that draws on the real instrument therefore has an unknown optimal statistic.  The best we can do in these cases is to form a noise model that includes all our knowledge about the noise, exactly as we do for the signal model.  Even when all distributions are known, the optimal statistic may be impractical to compute, especially when our signal and/or noise distributions are richly informative.  In either of these two scenarios, it is plausible that an \emph{ad hoc} statistic can be more practical.

However, we are not necessarily in that regime.  The statistics proposed in \cite{GuTi:89,Tinto96,KlMoRaMi:05,KlMoRaMi:06,MoRaKlMi:06,Ra:06} are special cases of a Bayesian model proposed in \cite{searle:08} which can be implemented at a comparable computational cost while using a more physical signal model. \cite{Ch_etal:06} explicitly considers noise bursts and attempts to reject them by computing an ``incoherent energy'' statistic, but the motivating ``bursty'' noise model can also be efficiently implemented in a Bayesian statistic, as outlined in \cite{searle:08}.

\section{Conclusion}

The conceptual simplicity of Monte-Carlo simulations have made them a popular tool to characterize the performance of gravitational wave burst searches.  They require the explicit adoption of a signal model in much the same fashion as a Bayesian analysis.  \emph{Ad hoc} proposals for statistics have not been optimal for the simulations they have been evaluated with; insofar as we trust these simulations to evaluate performance, we should consider, when practicable, going directly to the \emph{optimal statistics} they define in both the frequentist and Bayesian paradigms.  This trust can be better earned by using more physically plausible Monte-Carlo simulations: a physical continuous distribution of distances, orientations and inclinations instead of a handful of fixed distances, and a continuously parameterized and physically  motivated model for waveforms instead of a handful of injection templates.

\section*{Acknowledgments}
The author would like to thank Patrick Sutton, Massimo Tinto, Kipp Cannon, Alan Weinstein, Sergei Klimenko, Andrew Moylan and Zsuzsa Marka.  LIGO was constructed by the California Institute of Technology and Massachusetts Institute of Technology with funding from the National Science Foundation and operates under cooperative agreement PHY-0107417.  This document has been assigned LIGO Document Number LIGO-P080034-00-Z.
\section*{References}
\bibliography{main}
\end{document}